\documentclass[pra,showpacs,twocolumn]{revtex4}
\usepackage{bm,graphicx,amsmath}
\newcommand{\beq}{\begin{equation}}
\newcommand{\enq}{\end{equation}}
\begin{document}
\title{Ultracold atoms in a cavity mediated double-well system}
\author{Jonas Larson}
\email{jolarson@fysik.su.se}
\affiliation{Department of Physics, Stockholm University, Se-106 91 Stockholm, Sweden}
\author{Jani-Petri Martikainen}
\affiliation{NORDITA, Se-106 91
Stockholm, Sweden}

\date{\today}

\begin{abstract}
We study ground state properties and dynamics of a dilute ultracold atomic gas in a double
well potential. The Gaussian barrier separating the two wells
derives from the interaction between the atoms and a quantized field
of a driven Fabry-Perot cavity. Due to intrinsic atom-field
nonlinearity, several novel phenomena arise being the focus of this work. For the ground state, there is a critical pumping amplitude in which the atoms self-organize and the intra cavity field amplitude drastically increases. In the dynamical analysis, we show that the Josephson oscillations depend strongly on the atomic density and may be greatly suppressed within certain regimes, reminiscent of self-trapping of Bose-Einstein condensates in double-well setups. This pseudo self-trapping effect is studied within a
mean-field treatment valid for large atom numbers. For small numbers
of atoms, we consider the analogous many-body problem and demonstrate a
collapse-revival structure in the Josephson oscillations.
\end{abstract}
\maketitle

\section{Introduction}
Outcomes of nonlinearity have extensive consequences in various areas of physics leading to phenomena absent in their linear counterparts. In quantum mechanics it plays a crucial role in fields such as nonlinear optics~\cite{nonopt} and ultracold atomic gases~\cite{pethick}. In ultracold atomic gases, or Bose-Einstein condensates (BECs), the nonlinearity stems from interaction among the atoms, and cause effects as insulating states of cold atoms in optical lattices~\cite{mott}, soliton and vortex formation~\cite{soliton}, collapse-revivals of system evolution~\cite{milburn}, or self-trapping of atoms in optical lattices or double-well (DW) systems~\cite{smerzi}. Moreover, the inherent phase coherence of the BEC wave-function has made it a good candidate for realizing analogues of the Josephson effect appearing across the bulk of two attached superconductors~\cite{JJorig}. By placing the condensate in a DW potential, the weak tunneling through the center barrier brings about oscillations of the BEC between the two wells mimicking Josephson oscillation in superconductors~\cite{victor}. Nonlinearity in the BEC DW system induces self-trapping and collapse-revivals in the Josephson current. Self-trapping emerges for large nonlinearity and large imbalance of atoms between the two wells and manifests itself as blocking of the Josephson oscillations. Collapse-revivals characterize deaths and rebirths of the oscillations, and become important in the quantum regime where fluctuations around the mean-field condensate order parameter become relevant.  

In recent years, nonlinearity has turned out to be an essential ingredient in two different but highly linked matter-light quantum systems, optomechanical cavities~\cite{optom} and BEC-cavity setups~\cite{reichel,BECoptmech}. Here we focus on the second type in which pioneering experiments have demonstrated the square root dependence on the number of atoms $\Omega_{Rabi}\sim\sqrt{N}$ in the vacuum Rabi splitting~\cite{reichel}. Since then, most interest, both theoretically~\cite{teobist} and experimentally~\cite{BECbistable,BECoptmech}, has indeed been paid to the intrinsic nonlinearity within the system manifested, for example, by optical bistability. In these systems, the cavity field acts as an effective potential for the atoms, but the atoms,
in turn, induce a shift in the index of refraction altering the cavity field. As the field adjust accordingly, the field cause a back-action on the atoms resulting in the nonlinear atom-field interplay. Apart from studies of bistability, research has  considered also self-organization of ultracold atoms in optical resonators, either in setups of pumped cavities~\cite{selfcavity,jonas1} or pumped atoms~\cite{selfatom}, as well as most recently simulation of the Dicke quantum phase transition~\cite{dicke}. 

In this paper we focus on different outcomes of the atom-field nonlinearity compared to earlier works on optical bistability. That is, we are not mainly interested in multiple-solutions of the equations-of-motion, but instead direct our attention on self-organization, self-trapping, as well as collapse-revivals in a cavity mediated DW system. The appearance of additional solutions in nonlinear, in contrast to linear, systems is a quite general property. The present work therefore aims at more specific outcomes of atom-field nonlinearity. We notice that BEC DW systems coupled to cavity fields have been discussed previously~\cite{cavityDW,milburn2}, but those works, however, did not analyze the situation where the cavity field, and thereby the nonlinearity, drives the Josephson oscillations. Here self-organization and self-trapping are analyzed in the mean-field regime and our work goes beyond two-mode approximations commonly utilized in DW analysis. We especially show, as for the BEC DW, that the atomic density affects the strength of nonlinearity and thereby there is a critical atomic density where an effective self-trapping behavior appears. Self-organization emerges when the pump amplitude exceeds a critical value. At the same instant, the intra cavity field amplitude greatly increases similarly to the pump threshold in the theory of laser. Within the mean-field approach we also demonstrate how the system exhibits bistability as a result of nonlinearity. Collapse-revival structures derive from quantum fluctuations, and to tackle such effects we consider an effective two-mode many-body model for the system which indeed predicts such phenomenon. We also point out that the system setup automatically allows for quantum nondemolition measurements of the atomic dynamics via detection of the output cavity field.

\section{Josephson oscillations and the model system}
Before describing the system of the present work, we briefly summarize the
DW system in order to get a deeper understanding for the dynamics analyzed
later on in the paper.

\begin{figure}[ht]
\begin{center}
\includegraphics[width=7cm]{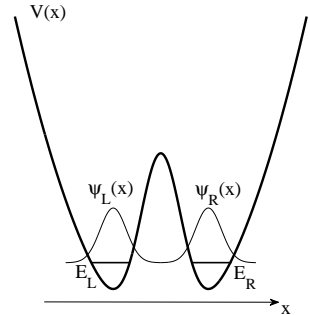}
\caption{Schematic picture of the traditional DW setup. In the
symmetric DW, the two onsite energies $E_L$ and $E_R$ are equal. The left
and right solutions are marked by $\psi_L(x)$ and $\psi_R(x)$
respectively.} \label{fig1}
\end{center}
\end{figure}

In the most simple situation, two wave-functions $\psi_L(x)$ and $\psi_R(x)$ are coupled with some strength $J$. In the symmetric situation, the groundstate and 
the excited state can be written as the symmetric and anti-symmetric
solutions
\begin{equation}\label{twomodeeq}
\begin{array}{l}
\Psi_0(x)=\frac{1}{\sqrt{\mathcal{N}}}\left(\psi_L(x)+\psi_R(x)\right),\\ \\
\Psi_1(x)=\frac{1}{\sqrt{\mathcal{N}}}\left(\psi_L(x)-\psi_R(x)\right),
\end{array}
\end{equation}
where $\mathcal{N}$ is the proper normalization coefficient. In DW-systems, $J$ represent the tunneling coefficient and $\psi_{L,R}(x)$ are the normalized left well and right well wave-functions as depicted in Fig.~\ref{fig1}. In general, we can express the time-dependent solution  
$\Psi(x,t)=\phi_L(t)\psi_L(x)+\phi_R(t)\psi_R(x)$ leading to a set
of two first order coupled equations for $\phi_L(t)$ and
$\phi_R(t)$. The corresponding Hamiltonian is
\begin{equation}\label{rabiham}
\hat{H}_{Rabi}=\left[\begin{array}{cc} E_L & -J \\
-J & E_R\end{array}\right],
\end{equation}
with $E_{L,R}$ the onsite energies of the left and right well,
and $J$ the tunneling strength. Thus, in the symmetric well, the detuning 
$\delta=E_L-E_R$ vanishes. The Rabi frequency, characterizing the Josephson oscillations, is given by $\Omega_{Rabi}=2\sqrt{J^2+\delta^2}/\hbar$,
which for the resonant well equals $2J/\hbar$. Nonzero detunings
imply an increased Josephson period, as well as causing the amplitude of the inversion
\begin{equation}\label{inversion}
\mathcal{Z}\equiv\frac{|\phi_R(t)|^2-|\phi_L(t)|^2}{N}=1-\frac{8J^2}{\delta^2+4J^2}\sin^2(\Omega_{Rabi}t)
\end{equation}
to decrease from oscillating between -1 and 1 as in the case of zero detuning. In
deriving (\ref{rabiham}), we have assumed  $\mathcal{Z}(t=0)=1$, i.e. all atoms initially in the right well.

\subsection{Cavity mediated double-well for cold atoms}
In both BEC DW experiments of Refs.~\cite{oberthaler}, the
potential barrier separating the left and right wells is obtained by
dispersive dipole interaction between the atoms and an external
laser beam. The laser has a transverse Gaussian mode shape, whose
width and amplitude are easily adjustable. The large laser intensity
makes the light field approximately classical. By confining atoms in
a resonator, the effective atom-field coupling can be greatly
enhanced and thereby the dynamics of single atoms can be affected by
the field even at average photon numbers less than
unity~\cite{kimble}. For such low intensities, the light field
sustained in the resonator must be treated quantum mechanically.

As we are dealing with a coupled bipartite quantum system, the state of the atoms will
influence the state of the field and vice versa. Physically, the
atomic matter wave brings out a change in the index of refraction,
and as the field adjust to the new index of refraction its change
will in return affect the atoms leading to an intrinsic nonlinear
atom-field interaction. For the present system it implies that the
effective DW potential directly depends on the atomic state.

\begin{figure}[ht]
\begin{center}
\includegraphics[width=6cm]{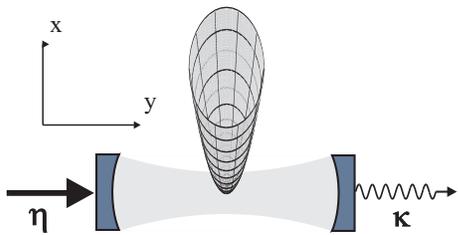}
\caption{(Color online) System setup. An anisotropic harmonic trap confines the
atoms such that the motion can be considered quasi one-dimensional
along the $x$-direction. The field of a Fabry-Perot cavity, aligned
along the $y$-direction, intersects the atomic trap. The $TEM_{00}$-mode shape
of the cavity induces an effective Gaussian barrier separating the
harmonic trap into two wells. Furthermore, the cavity is laser-driven through one end mirror with an amplitude $\eta$. Cavity losses, with decay rate $\kappa$, are marked by a curly arrow. } \label{fig2}
\end{center}
\end{figure}

The system we have in mind is illustrated in
Fig.~\ref{fig2}. A harmonic potential traps the atoms in all three
dimensions, but the trap frequencies in the $y$- and $z$-directions
are assumed large enough so that the atoms remain in their lowest
vibrational states in these directions, and we consequently consider
motion restricted to the $x$-direction. Instead of an external
laser, we consider the field of a cavity to constitute the tunneling
barrier. To this end we use a Fabry-Perot cavity with its
longitudinal axis along the $y$-direction. As for a traditional
laser beam, the Fabry-Perot cavity possesses $TEM_{00}$-modes having transverse
Gaussian profiles. The cavity has a large enough $Q$-value to
guarantee well separated mode frequencies. Nonetheless,
cavity losses $\kappa$ are taken into account in all our derivations. Only
a single cavity mode is quasi resonant with the atomic transition
under consideration, and all other modes are therefore ignored. The
cavity is externally driven with a classical field having amplitude
$\eta$, and hence, without atoms present the cavity field would
reach a steady state being coherent with an amplitude determined by
the balancing of losses $\kappa$ and pumping $\eta$. The cavity
is taken to couple dispersively to the atoms and the excited atomic state can
hence be eliminated adiabatically~\cite{jonasad}. In a frame
rotating with the pump frequency $\omega_p$ and assuming ultracold
atoms, the Hamiltonian reads~\cite{jonas1}
\begin{equation}\label{2ham}
\begin{array}{lll}
\hat{H}_{af} & = & -\hbar\Delta_c\hat{a}^\dagger\hat{a}-i\hbar\eta\left(\hat{a}-\hat{a}^\dagger\right)\\ \\
& & \displaystyle{+\int dx\hat{\Psi}^\dagger(x)\left[-\frac{\hbar^2}{2m}\frac{\partial^2}{\partial x^2}+\frac{m\omega^2x^2}{2}\right.}\\ \\
& & \displaystyle{\left.+ U(x)\hat{a}^\dagger\hat{a}+\frac{gN}{2}|\hat{\Psi}(x)|^2\!\right]\hat{\Psi}(x),}
\end{array}
\end{equation}
where $\Delta_c=\omega_p-\omega_c$ is the pump-cavity detuning,
$\hat{a}^\dagger$ ($\hat{a}$) the photon creation (annihilation)
operator obeying standard boson commutation rules, $\omega$ the trap
frequency, and
\begin{equation}
U(x)=\hbar U_0e^{-\frac{x^2}{\Delta_x^2}}
\end{equation}
is the effective dispersive atom-field coupling with
$U_0=\lambda^2/\Delta_a$ where $\lambda$ is the single photon
atom-field coupling, $\Delta_a=\omega_p-\omega_a$ the pump-atom
detuning, and $\Delta_x$ is the mode waist. Note that in order to achieve a DW structure we 
restrict the analysis to positive detunings, $\Delta_a>0$, but interesting effects
can also appear for negative detunings when the cavity field induces a potential
``dimple'' on the atoms and consequently raises the local phase space density.
In this work, choosing $U(x)$ centered around $x=0$ we consider the symmetric DW, and consequently the $dc$ (direct current) Josephson effect. The $ac$ (alternating current) Josephson effect is assessed by spatially shifting either the cavity or the trapping potential. The last nonlinear term on the right hand side of Eq.~(\ref{2ham}) stems from atom-atom collisions and is generally proportional to the atomic density and the $s$-wave scattering length. 

After introducing field losses, $\kappa$, the corresponding Heisenberg-Langevin equations become
\begin{equation}
\begin{array}{lll}
\displaystyle{\frac{d}{dt}\hat{a}} & = & \displaystyle{-(\kappa-i\Delta_c)\hat{a}+\eta}\\ \\
& & \displaystyle{-iN\int|\hat{\Psi}(x)|^2U(x)\,dx\hat{a}+\sqrt{2\kappa}\hat{a}_{in}(t)},\\ \\
\displaystyle{\frac{d}{dt}\hat{\Psi}(x)} & = & \displaystyle{\Bigg[-\frac{\hbar^2}{2m}\frac{\partial^2}{\partial x^2}+\frac{m\omega^2x^2}{2}}\\ \\
& & \displaystyle{+U(x)\hat{a}^\dagger\hat{a}+gN|\hat{\Psi}(x)|^2\Bigg]\hat{\Psi}(x).}
\end{array}
\end{equation}
Here, $\hat{a}_{in}(t)$ is the Langevin field
input noise source being $\delta$-correlated;
$\langle\hat{a}_{in}(t)\hat{a}_{in}(t')^\dagger\rangle=\delta(t-t')$,
and
$\langle\hat{a}_{in}^\dagger(t)\hat{a}_{in}(t')\rangle=\langle\hat{a}_{in}(t)\rangle=0$ otherwise~\cite{carmichael}.
In the present work, the influence of quantum noise will be assumed
small and hereafter neglected. The effects of such fluctuations in a
BEC cavity system similar to the one we study has been analyzed in
Ref.~\cite{domokosnoise}.

In typical experiments, the characteristic time-scales for the field
and atoms are substantially different~\cite{reichel}. The field
evolution can be assumed to follow the dynamics of the atoms and it
is therefore justified to consider the steady state solutions of the field.
Explicitly, the steady state for the photon number
$\hat{n}=\hat{a}^\dagger\hat{a}$ is given by
\begin{equation}\label{ssphoton}
\hat{n}_{ss}=\frac{\eta^2}{\kappa^2+\left(\Delta_c-N\hat{Y}\right)^2},
\end{equation}
with the operator $\hat{Y}=\int|\hat{\Psi}(x)|^2U(x)\,dx$. 
This equations makes clear the modification of the detuning $\Delta_c$ induced by 
the atoms, i.e. the resonance condition $\Delta_c=0$ is shifted to $\Delta_c=N\hat{Y}$. Since the
atomic field $\hat{\Psi}(x)$ is coupled to the atom number
$\hat{n}$, it follows that so is $\hat{Y}$ and then
Eq.~(\ref{ssphoton}) may render multiple solutions of the photon
number. This same nonlinear effect give rise to quantum optical
bistability~\cite{teobist,BECbistable} which can furthermore be shown to
be analogous to bistable optomechanics in certain regimes~\cite{BECoptmech}.

The intrinsic atom-field nonlinearity is rather different from the
nonlinearity originating from atom-atom interactions~\cite{duncan}.
The number of photons enters in the effective system potential, and
since it depends on the matter state the effective potential can be
seen as a dynamical variable. In the BEC DW situation, nonlinearity enters in the atom-atom interaction term, and within the Thomas-Fermi regime (where kinetic energies can be neglected) it can be viewed as changing the chemical potentials in the two wells. Roughly speaking, as an outcome in the BEC DW the effective shift due to nonlinearity enters in the detuning $\delta$, while in the cavity DW it modifies field strength $\hat{n}_{ss}$ and thereby the tunneling coefficient $J$.

Throughout, we try to use
realistic experimental parameters. More explicitly, we consider the
cavity decay rate $\kappa$ from the experiment of the Esslinger
group~\cite{BECbistable}, and express other rates in terms of this.
Lengthscales as well as atom numbers are taken as to be
experimentally realistic.

\section{Stationary solutions}
In this and the next section we discuss the system at a mean-field level, i.e. replacing the atomic operator $\hat{\Psi}(x)$ by its mean. The resulting Gross-Pitaevskii equation~\cite{ritsch}   
\begin{equation}\label{GPeq}
\begin{array}{lll}
i\displaystyle{\frac{d}{dt}\Psi(x)} & = & \displaystyle{\Bigg[-\frac{\hbar^2}{2m}\frac{\partial^2}{\partial x^2}+\frac{m\omega^2x^2}{2}}\\ \\
& & \displaystyle{+n_{ss}
U(x)+gN|\Psi(x)|^2\Bigg]\Psi(x),}
\end{array}
\end{equation}
for the order parameter can be solved for the minimum energy solution by propagating it
in imaginary time and updating the cavity photon number 
\begin{equation}\label{ssphotonmf}
n_{ss}=\frac{\eta^2}{\kappa^2+(\Delta_c-NY)^2}
\end{equation}
during the propagation so that it corresponds
to the instantaneous steady-state photon number. Here $Y$ is the mean-field counterpart of the operator $\hat{Y}$.
The corresponding energy functional is
given by~\cite{duncan}
\begin{equation}
\begin{array}{lll}
\displaystyle{\frac{E[\Psi]}{N}} & = & \displaystyle{\int\left[\left(\frac{\hbar^2}{2m}\right)\left|\frac{\partial\Psi(x)}{\partial x}\right|^2\right.}\\ \\ & & \displaystyle{+\left.\frac{m\omega^2x^2}{2}|\Psi(x)|^2+\frac{gN}{2}|\Psi(x)|^4\right]dx}  \\ \\
& & \displaystyle{-\frac{\eta^2}{\kappa N}\arctan\left(\frac{\Delta_c-N\int U(x)|\Psi(x)|dx}{\kappa}\right).}
\end{array}
\end{equation}
In this paper we are not interested in the interaction effects due to atomic collisions, but
wish to focus on the coupling between the cavity mode and the atoms.
Therefore, we chose $g=0$ which can be achieved via Feshbach
resonances~\cite{feshbach} or is approximately valid in a sufficiently dilute gas.

While a direct solution of the Gross-Pitaevskii equation is relatively easily 
found numerically, we found that
in many instances most of the equilibrium physics (and some of the dynamics as well)
can also be captured by a more transparent variational ansatz.
Our ansatz is a Gaussian ansatz which can have two peaks, namely
\begin{equation}\label{doublepeakedgauss}
\psi(x)=C\left[e^{-\frac{(x+x_0)^2}{2\sigma^2}}+e^{-\frac{(x-x_0)^2}{2\sigma^2}}\right],
\end{equation}
where the prefactor is set by the normalization $\int dx |\psi(x)|^2=1$ and is given by
\beq
C=\frac{1}{\pi^{1/4}\sqrt{2\sigma^2(1+\exp\left[-\left(\frac{x_0}{\sigma}\right)^2\right])}}.
\enq
Substituting this ansatz into the energy functional gives us an energy functional
$E(\sigma,x_0)$ which can be minimized to find the optimal solution. All integrals
are sufficiently easy to solve analytically, but the resulting expressions are too 
long to be given here explicitly.

The steady state solutions naturally separate into two different regimes, one at
the low values of the pumping strength and the other at large pumping strengths.
When the pumping strength is small the atomic order parameter
is nearly a Gaussian centered around the origin. In this limit the atomic density overlaps
strongly with the Gaussian cavity mode function and if the coupling strength $U_0$
is quite large, $Y$ can be substantial. 
Then the steady state
photon number can be suppressed by the $NY$-term appearing in the denominator 
of Eq.~(\ref{ssphotonmf}) if
$|NY|\gg \Delta_c$. As the pumping is increased, the photon number increases at first
roughly $\propto \eta^2$ followed by a pronounced jump in the cavity photon number at some
critical pump strength.

This jump coincides with a qualitative change in the atomic order parameter as the 
atomic density splits into two separate peaks and develops a minimum at the origin.
This transition is signaled by a rapid increase in the cavity photon number since
the overlap of the atomic density and the cavity mode function is rapidly
reduced so that the $NY$-contribution in the steady-state photon number becomes 
less important. At the same time the effective potential experienced by the atoms
becomes a stronger double well potential, which pushes the atomic
density peaks further apart and lowers the overlap with the cavity mode function even more.
In a way, as one increases the pumping, one can move from a regime where
coupling between the atomic order parameter and cavity field is strong into a regime
where it is weak and the cavity field mainly acts as an independent optical potential
acting on the atoms. 

This drastic change in the atomic density of the groundstate is an example of self-organization~\cite{selfatom,selforg}. If the atoms are externally pumped, rather than the cavity field, the atoms act as a scatter which transfers photons from the pump into the resonator. Thereby, there is a direct resemblance between pumping of the atoms and pumping of the cavity field. The corresponding phenomenon for pumped atomic systems has been studied in great detail and it was found that the transition is described by a second order phase transition~\cite{selforg2}. The same behavior reminds of threshold-pumping of lasers. Pumping of the laser must exceed a critical value in order for lasing to set off. Again, such a critical structure has been identified as a second-order phase transition~\cite{laser}. Moreover, it is worth pointing out that in a very recent experiment, the same kind of transition has been demonstrated for a driven lossy cavity in circuit QED~\cite{laser2}.

The critical pump strength which separates these two self-organized regimes
occurs roughly when the barrier height becomes larger than the energy scale of the
background harmonic oscillator i.e. when $n_{ss}\hbar U_0\sim \hbar\omega$.
The resonance in the photon number occurs when $\Delta_c=NY$. For the Gaussian groundstate
of the non-interacting system (which is accurate at small pump strengths) this implies
$NU_0/\Delta_c=\sqrt{1+\sigma/\Delta_x^2}$, where $\sigma=\sqrt{\hbar/m\omega}$
is the width of the non-interacting (Gaussian) atomic wavefunction. 
In this paper we mostly use $\Delta_x=\sigma/2$.

\begin{figure}[ht]
\begin{center}
\includegraphics[width=8cm]{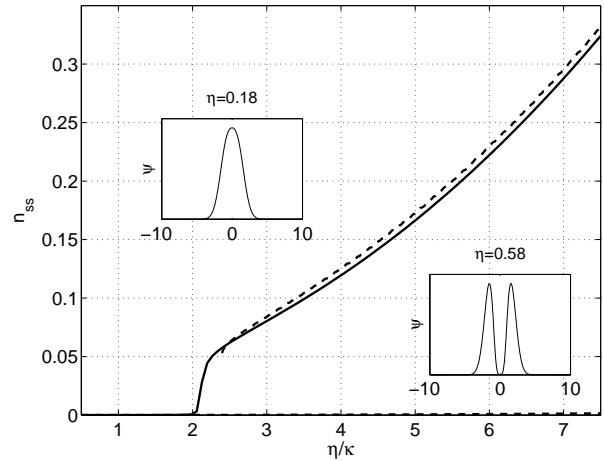}
\caption{The mean-field photon number as a function of $\eta/\kappa$
using a double-peaked gaussian ansatz (dashed-line) and Gross-Pitaevskii
approach (solid line). With some parameters, the Gaussian ansatz predicts
an existence of two stationary solutions corresponding to local minima.
Both of these solutions are indicated in the figure. As insets we show 
the wavefunction amplitude $|\Psi(x,t)|$ using the Gross-Pitaevskii equation both at smaller and higher
pumping strength.
We used parameters $N=10000$, $\kappa=2\pi\times1.3$
MHz, $\Delta_c=\kappa$, $U_0=\kappa/100$, 
$\Delta_x=0.5\sqrt{\hbar/m\omega}$, $\omega=\kappa/500$, and $m=87u$.}
\label{Fig:GaussianvsGP}
\end{center}
\end{figure}

In Fig.~\ref{Fig:GaussianvsGP} we show an example of the behavior of the steady-state photon number as a function of pumping together with few examples of the associated 
atomic order parameters as obtained from imaginary time propagation of the Gross-Pitaevskii equation. In this figure, we also compare the double peaked
Gaussian ansatz with the results from the Gross-Pitaevskii equation and find the 
agreement to be very good. Using the Gaussian ansatz
we can, depending on parameters, find two stationary solutions corresponding to 
minima of the energy functional. The upper branch is the global minimum, while the lower one
is locally stable. Our Gross-Pitaevskii approach only finds the global minimum and in order
to find the other solution, we would have to impose additional constraints on the imaginary time propagation.

In agreement with earlier works on atom-light
bistability~\cite{teobist,BECbistable}, the present model
supports two or one stable values of $n_{ss}$. In addition to these
solution there exists the ``middle'' branch of the hysteresis curve representing
an unstable solution corresponding to a local energetic maximum.
We have not plotted this branch in Fig.~\ref{Fig:GaussianvsGP}.

\begin{figure}[ht]
\begin{center}
\includegraphics[width=8cm]{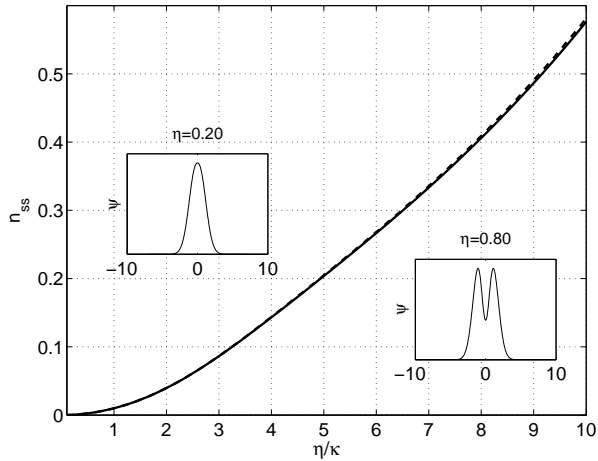}
\caption{$n_{ss}$
using a double-peaked gaussian ansatz (dashed-line) and Gross-Pitaevskii
approach (solid line).  As insets we show 
the wavefunction amplitude $|\Psi(x,t)|$ using the Gross-Pitaevskii equation both at smaller and higher
pumping strength.
We used parameters $N=10000$, $\kappa=2\pi\times1.3$
MHz, $\Delta_c=\kappa$, $U_0=\sqrt{5}\kappa/N\approx 0.000224 \,\kappa$, 
$\Delta_x=0.5\sqrt{\hbar/m\omega}$, $\omega=\kappa/500$, and $m=87u$.}
\label{Fig:Gaussian_smallU0}
\end{center}
\end{figure}

The above discussion is based on a fairly high value ($U_0=\kappa/100$)
of atom-field coupling. For a lower value of $U_0$ the behavior can be quite different. As an example,
we show in Fig.~\ref{Fig:Gaussian_smallU0} results using the double peaked Gaussian ansatz 
and the Gross-Pitaevskii approach
with otherwise the same parameters, but a much smaller coupling $U_0\approx 0.000224 \kappa$. We chose this value
because then $\Delta_c-NY=0$ and the steady state photon number is on resonance 
for the ideal gas Gaussian groundstate, when other parameters
are kept the same as in Fig.~\ref{Fig:GaussianvsGP}. In this case the large overlap between
the atomic order parameter and the cavity mode function acts as to enhance the steady state
photon number. This is in contrast to the behavior at larger couplings where similar large overlap
implied strong suppression of the cavity photon number. In this case we can find no evidence 
of bistability and the double peaked atomic order parameter appears smoothly as the pumping strength
is increased. At the same time the overlap between the cavity mode function and the atomic
density decreases smoothly.

\section{Mean-field dynamics}
In this section we give an overview of the typical features
of the system dynamics at the mean-field level. Precise details
naturally vary based on the used parameters, but the basic structures
and the physical picture behind them is robust with respect to
changes in the parameters.

\begin{figure}[ht]
\begin{center}
\includegraphics[width=8cm]{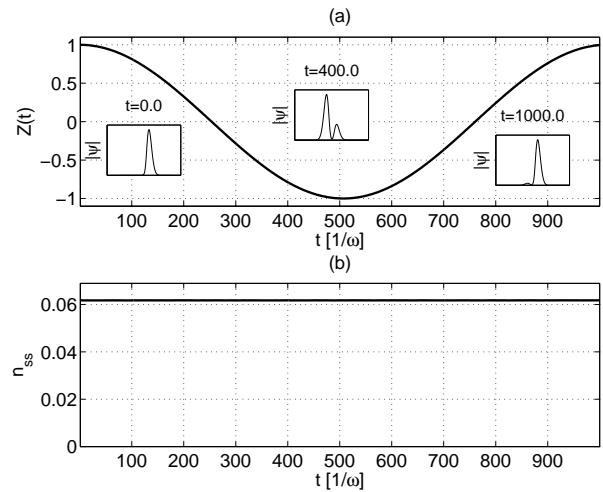}
\caption{Dynamics of the atomic order parameter when the initial state
was localized to the right of the cavity mode barrier. The upper figure (a) shows
the inversion $\mathcal{Z}(t)$ together
with few snapshots of the absolute value $|\Psi(x,t)|$ of the atomic order parameter. The lower plot (b) displays the photon number $n_{ss}$. We used parameters $N=10000$, $\kappa=2\pi\times1.3$
MHz, $\eta=25\, \kappa$, $\Delta_c=\kappa$, $U_0=\kappa/200$, 
$\Delta_x=0.5\sqrt{\hbar/m\omega}$, $\omega=\kappa/500$, and $m=87u$.}
\label{Fig:NiceRabi}
\end{center}
\end{figure}

We can construct wavepackets which are localized either left or right of the 
barrier by finding the symmetric ($\psi_{sym}(x)$)
and anti-symmetric solutions ($\psi_{asym}(x)$) to the Gross-Pitaevskii
equation. Using these solutions 
the orthogonal solutions localized either to the left or right
are given by
\beq
\psi_L(x)=(\psi_{sym}(x)-\psi_{asym}(x))/\sqrt{2}
\enq
and
\beq
\psi_R(x)=(\psi_{sym}(x)+\psi_{asym}(x))/\sqrt{2}.
\enq
In Fig.~\ref{Fig:NiceRabi} we present an example of the dynamics for an initial state which is
localized to the right of the barrier. Since the barrier height in this case is quite high, the overlap
between the mode function and the order parameter is small. This remains true even in the
course of the dynamics and the cavity photon number has only  weak time-dependence.
The atomic order parameter then exhibits Rabi-flopping from the right localized state
to the left localized state with a period which becomes longer as the barrier gets higher
and/or the cavity mode function wider. Plotted is the inversion
\begin{equation}
\mathcal{Z}(t)=1-2\int_{-\infty}^0\,|\Psi(x,t)|^2dx,
\end{equation}
and the photon number~(\ref{ssphotonmf}). For the utilized parameters, $t=100$ in dimensionless variables corresponds to approximately 6 ms. 

\begin{figure}[ht]
\begin{center}
\includegraphics[width=8cm]{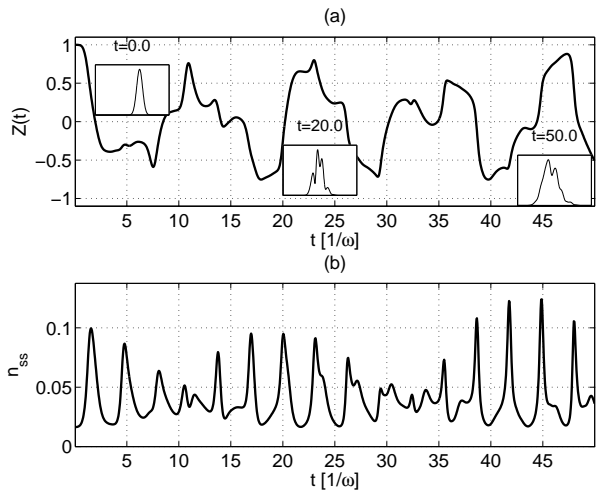}
\caption{Dynamics of the atomic order parameter when the initial state
was a Gaussian localized at $x_0=2$ to the right of the cavity mode barrier and having a width $\sigma=0.8$. 
The upper figure (a) shows
the inversion $\mathcal{Z}(t)$ together
with few snapshots of the absolute value $|\Psi(x,t)|$ of the atomic order parameter, while the lower plot gives the photon number $n_{ss}$.
We used parameters $N=10000$, $\kappa=2\pi\times1.3$
MHz, $\eta=40\, \kappa$, $\Delta_c=3\kappa$, $U_0=\Delta_c\sqrt{5}/N$, 
$\Delta_x=0.5\sqrt{\hbar/m\omega}$, $\omega=\kappa/500$, and $m=87u$.}
\label{Fig:MultimodeDyn}
\end{center}
\end{figure}

In Fig.~\ref{Fig:MultimodeDyn} we show another example of the dynamics for an initial state which is localized to the right of the barrier, but with a smaller $U_0$ and larger $\Delta_c$. In this case, the cavity photon number does have pronounced time-dependence and the system cannot be described as a simple Rabi-flopping anymore. The reason for the dramatically different dynamics is that with the parameters used in Fig.~\ref{Fig:MultimodeDyn} the initial barrier height is quite low and proportional to $1/(\kappa^2+\Delta_c^2)$ since the overlap between the atomic density and the cavity mode is low. This low barrier height enables the wavefunction to populate also the barrier region to a greater extent, which increases the overlap with the cavity mode. However, such increase lowers the $(\Delta_c-NY)^2$ term in the denominator of Eq.~(\ref{ssphotonmf}) and causes the photon number to increases toward its maximum value $\propto 1/\kappa^2$. This increase in turn increases the barrier height which tends to drive the atomic order parameter away from the region close to the origin. A final result of this complicated interplay is a pronounced correlated dynamics of the atoms and the cavity field. The cavity field reaches a maximum when atoms are, on the average, more closely located to the center and is a minimum when the order parameter has only a small overlap with the cavity mode.

It should be noticed that while the results in Fig.~\ref{Fig:NiceRabi} can be analyzed
using a two-mode description~\cite{smerzi} the results  presented in Fig.~\ref{Fig:MultimodeDyn} cannot be analyzed in that way and a multi-mode description is essential. Since the field amplitude is proportional to the
photon number, detection of the output cavity field intensity would
directly reveal some properties of the atom dynamics in the cavity. Moreover, such
recording of the output cavity field is non-destructive. The idea of
utilizing dispersive cavity interaction for non-demolition
measurements has been discussed in terms of BEC DW
systems~\cite{milburn2}, or for cold atoms trapped in optical
lattices~\cite{ritschQND}. These references, however, do not
consider the cavity field as supplying an effective potential and
thus is very different from the present system.

\begin{figure}[ht]
\begin{center}
\includegraphics[width=8cm]{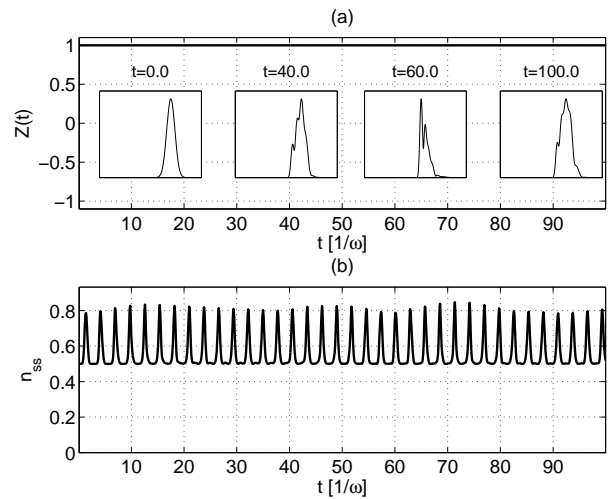}
\caption{Dynamics of the atomic order parameter when the initial state
was an excited state with a Gaussian profile 
localized to the right of the cavity mode barrier. The upper figure (a) shows
the inversion $\mathcal{Z}(t)$ together
with few snapshots of the absolute value $|\Psi(x,t)|$ of the atomic order parameter. Within the time period of this plot, the atoms are localized within the right well, i.e. pseudo self-trapped. The lower one displays the photon number $n_{ss}$. 
We used parameters $N=10000$, $\kappa=2\pi\times1.3$
MHz, $\eta=100\kappa$, $\Delta_c=\kappa$, $U_0=\kappa/200$, 
$\Delta_x=0.5\sqrt{\hbar/m\omega}$, $\omega=\kappa/500$, and $m=87u$.}
\label{Fig:EffectiveLoc}
\end{center}
\end{figure}

When the system was prepared in a localized state whose energy is close to the 
ground state of the double well potential, we found a simple
Rabi flopping behavior in Fig.~\ref{Fig:NiceRabi}.
In Fig.~\ref{Fig:EffectiveLoc} we demonstrate how the state can become 
effectively localized to the one side of the double well system when
it is prepared in a localized excited state. Here we prepared the atoms
in a displaced Gaussian state and let the system evolve with somewhat higher
cavity pumping strength than in Fig.~\ref{Fig:NiceRabi}. As the
atoms approach the barrier region the photon numbers rise, make the 
barrier higher, and tend to push the atoms back. In this case 
the atoms effectively stay localized in one well and 
the cavity photon number has a regular variation which reflects the average 
position of the atoms in a DW system. In this respect, the localization is reminiscent of self-trapping. In regular BEC DW systems, increasing the atom number implies enhanced nonlinearity and a transition between delocalized to localized states. This pseudo self-trapping is not perfect; the tunneling rate becomes very small but it is still finite, while in BEC DW systems it is rather the detuning that is affected by the nonlinearity and therefore the oscillations are never perfect in this case. Note, however, that within typical experimental time scales one would expect well established localization.

\begin{figure}[ht]
\begin{center}
\includegraphics[width=8cm]{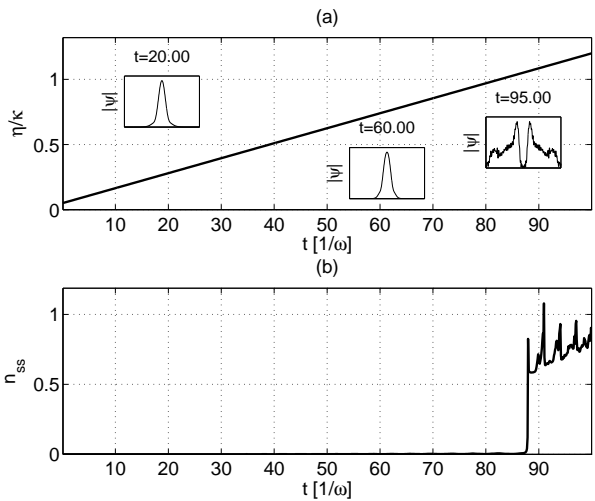}
\caption{The pump strength (a), mean-field photon number (b), and snapshots (insets) of 
the atomic order parameter $|\Psi(x,t)|$ as a function of time when the pump strength is sweeped from
low to high values. After a certain pump strength, the barrier separating the left and right wells becomes sufficiently large to split the atom density and simultaneously the photon number makes a drastic change. We used parameters $N=10000$, $\kappa=2\pi\times1.3$
MHz, $\Delta_c=\kappa$, $U_0=\kappa/200$, 
$\Delta_x=0.5\sqrt{\hbar/m\omega}$, $\omega=\kappa/500$, and $m=87u$.}
\label{Fig:SweepEtaUp}
\end{center}
\end{figure}

\begin{figure}[ht]
\begin{center}
\includegraphics[width=8cm]{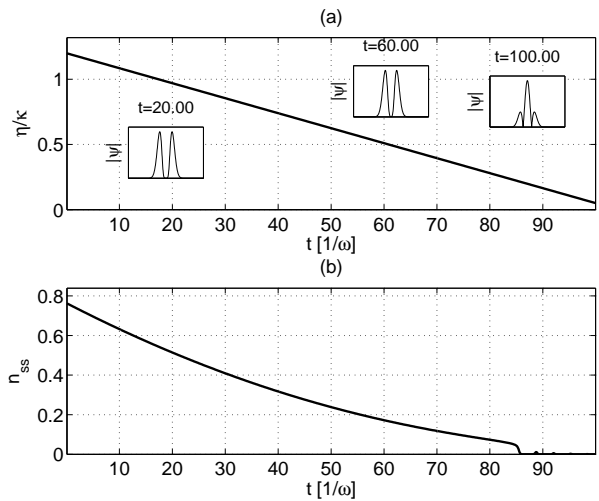}
\caption{The pump strength (a), mean-field photon number (b), and snapshots (insets) of 
the atomic order parameter $|\Psi(x,t)|$ as a function of time when the pump strength is sweeped from
high to low values. In comparison to Fig.~\ref{Fig:SweepEtaUp}, the transition is here much smoother. We used the same parameters as in Fig.~\ref{Fig:SweepEtaUp} i.e. $N=10000$, $\kappa=2\pi\times1.3$
MHz, $\Delta_c=\kappa$, $U_0=\kappa/200$, 
$\Delta_x=0.5\sqrt{\hbar/m\omega}$, $\omega=\kappa/500$, and $m=87u$.}
\label{Fig:SweepEtaDown}
\end{center}
\end{figure}

In finding the steady state solution to the Gross-Pitaevskii equation using
the imaginary time propagation, we converged to the state corresponding
to the upper branch of the Gaussian ansatz and ignored the other stationary state.
However, even if this state is not the lowest energy state it can play an important role
in the dynamical behavior of the atoms. We illustrate this with few examples in 
Figs. ~\ref{Fig:SweepEtaUp} and ~\ref{Fig:SweepEtaDown}. In Fig. ~\ref{Fig:SweepEtaUp} we start from the steady state at small pumping strengths and then ramp it up while
in Fig. ~\ref{Fig:SweepEtaDown} we do the reverse. Comparing the results indicates a
hysteresis type of behavior. 

When the pumping field is ramped up, at first very little happens either
for the atomic order parameter or the cavity field. There is a broad range of pumping strength above the critical pumping strength for the transition to a doubled peaked atomic density distribution where the system is still dynamically, but not thermodynamically, stable. Only when the pumping strength exceeds some higher threshold does the system become dynamically unstable. When this happens, the atomic order parameter quickly splits, the cavity photon number is suddenly increased, and the energy of the atoms is increased considerably in this abrupt process. 

This behavior is easy to understand, since here we start with a single peaked
atomic order parameter which has a substantial overlap with the cavity mode function. This means that the photon numbers, and hence the barrier height, are strongly suppressed by the 
$NY$-term in the denominator. Only once the pumping strength becomes so strong that the
barrier height is increased sufficiently to split the wavefunction, does the picture change.
Then the overlap parameter $Y$ is quickly reduced and the photon number (and the barrier height) increase. This in turn splits the atomic order parameter even more and results in the abrupt dynamics we observe.

On the other hand, when the field is ramped down (Fig. ~\ref{Fig:SweepEtaDown}) the
initial order parameter has only small overlap with the cavity mode function. Reducing
the pumping strength will then reduce the barrier height smoothly allowing the order
parameter to contracts toward the single peaked order parameter more smoothly. Eventually
the overlap between the order parameter and the cavity mode becomes so strong that
the $NY$-term in the denominator of Eq. (\ref{ssphotonmf}) reduces the photon number more strongly than the simple reduction of the pumping strength would suggest.

Up till now we considered an atom number $N=100$ which is of the correct order of magnitude with present experiments~\cite{reichel}. The thermodynamical limit is given by letting $N$ and the effective mode volume $V$ tend to infinity while keeping the density $\rho=N/V$ fixed. In the driven cavity QED system one has $U_0\propto V^{-1}$ and $\eta\propto\sqrt{V}$~\cite{cavityQED}, and it follows that the effective potential $n_{ss}U(x)$ depends solely on $\rho$, not $N$ and $V$ independently. Consequently, the same characteristics is obtainable for other atom numbers $N$ by suitably choosing $eta$ and $U_0$.

\section{Dynamics beyond mean-field}
Since it is known that often the relevant physics can be captured
by a two-mode model we consider such an approach in this section which in particular allows us to explore physics beyond mean-field. Especially for deep enough DWs, it is legitimate to assume that the ground state can be written as in Eq.~(\ref{twomodeeq}), i.e.
$\Psi_0(x)=\frac{1}{\sqrt{2}}\left(\psi_L(x)+\psi_R(x)\right)$ with
$\psi_{L,R}(x)$ localized states in the left and right wells. As in the previous sections, the left and right functions $\psi_{L,R}(x)$ are taken to be Gaussian in form. We pick the center of these to coincide with the minima of the effective potential
\begin{equation}
V_{eff}(x)=\frac{m\omega^2x^2}{2}+U(x)n_{ss},
\end{equation}
giving
\begin{equation}
x_0=\Delta_x\sqrt{\ln\left(\frac{2\hbar U_0n_{ss}}{\Delta_x^2m\omega^2}\right)}.
\end{equation}
The width, on the other hand, is obtained from minimizing the corresponding energy functional. In principle, $x_0$ could be kept a variational parameter, but in order
to simplify the estimations of the parameters of our two-mode model, we 
impose this additional assumption. Relaxing this constraint, is straightforward, but is not expected to change our results in an important way, especially
when the barrier height is large compared to the energy scale of the harmonic trap.

Thus, the width $\sigma$ is found by minimizing the system's energy
functional. This results in a set of coupled nonlinear equations for
$\sigma$ and $n_{ss}$ which must be solved self-consistently. In
particular, multiple solutions of the equations may exist in certain
parameter regimes. Introducing the overlap integrals
\begin{equation}
\begin{array}{l}
\displaystyle{E_0=-\frac{\hbar^2}{2m}\int\psi_i^*(x)\left(\frac{d^2}{dx^2}\right)\psi_i(x)dx,}\\ \\
\displaystyle{E_1=-\frac{\hbar^2}{2m}\int\psi_L^*(x)\left(\frac{d^2}{dx^2}\right)\psi_R(x)dx,}\\ \\
\displaystyle{J_0=\int|\psi_i(x)|^2e^{-x^2/\Delta_x^2}dx},
\\ \\
\displaystyle{J_1=\int\psi_L^*(x)\psi_R(x)e^{-x^2/\Delta_x^2}dx},
\\ \\
\displaystyle{S_0=\frac{m\omega^2}{2}\int|\psi_i(x)|^2x^2dx},
\\ \\
\displaystyle{S_1=\frac{m\omega^2}{2}\int\psi_L^*(x)\psi_R(x)x^2dx},
\end{array}
\end{equation}
the groundstate energy within the present ansatz takes the form~\cite{duncan}
\begin{equation}\label{energyfun}
\begin{array}{lll}
\displaystyle{\frac{E}{N}} & = & E_0+E_1+S_0+S_1\\ \\
& & \displaystyle{-\frac{\eta^2}{\kappa N}\arctan\left(\frac{\Delta_c-U_0N(J_0+J_1)}{\kappa}\right).}
\end{array}
\end{equation}
We artificially impose orthogonality of the left and right well wave-functions
($\int \psi_L^*(x)\psi_R(x)dx=0$) in order to avoid
unphysical contributions, and then get
\begin{equation}\label{coeff2}
\begin{array}{lll}
\displaystyle{E_0=\frac{\hbar^2}{4m\sigma^2}},& &
\displaystyle{E_1=\frac{\hbar^2}{4m\sigma^2}e^{-\frac{x_0^2}{\sigma^2}}}\\ \\
\displaystyle{J_0=\frac{\Delta_x}{\sqrt{\Delta_x^2+\sigma^2}}e^{-\frac{x_0^2}{\Delta_x^2+\sigma^2}}},
& &
\displaystyle{J_1=\frac{\Delta_x}{\sqrt{\Delta_x^2+\sigma^2}}e^{-\frac{x_0^2}{\sigma^2}}},
\\ \\
\displaystyle{S_0=\frac{m\omega^2}{2}\left(x_0^2+\frac{\sigma^2}{2}\right)},
& &
\displaystyle{S_1=\frac{m\omega^2}{2}\frac{\sigma^2}{2}e^{-\frac{x_0^2}{\sigma^2}}}.
\end{array}
\end{equation}
As pointed out, in order to analyze effects beyond mean-field we assume the
system parameters to be such that we can impose a two-mode
approximation. Thus, we expand the atomic operators as 
$\hat{\Psi}(x)=\hat{b}_L\psi_L(x)+\hat{b}_R\psi_R(x)$, where
$\hat{b}_{L,R}$ ($\hat{b}_{L,R}^\dagger$) annihilates (creates) an
atom in well $L,\,R$, and $\psi_{L,R}(x)$ are determined as above. The two functions $\psi_{L,R}(x)$ are
operator valued since they depend on the cavity field amplitude. As
an outcome, in deriving equations-of-motion or an effective
Hamiltonian for the atoms when the cavity field is eliminated, one
has to take ordering between non-commuting operators into
account~\cite{jonas1}. To do so, we will assume $J_0\gg J_1$ and
leave out the $J_1$ cross-term in $\hat{Y}$ for the steady-state
photon number of Eq.~(\ref{ssphoton}). For the present potential
$U(x)$, this is not always justified but nevertheless it holds in
large parameter regimes. In this work we will not present the full
derivation of the effective Hamiltonian, but refer to
Ref.~\cite{jonas1} for details. To order $1/N^2$ we find a second
quantized Hamiltonian for the atoms
\begin{equation}
\hat{H}_{BH}=(E_0+S_0)\hat{N}+f(\hat{N})-t(\hat{N})\hat{B},
\end{equation}
where
\begin{equation}
\begin{array}{l}
\hat{N}=\hat{n}_R+\hat{n}_L=\hat{b}_R^\dagger\hat{b}_R+\hat{b}_L^\dagger\hat{b}_L,\\ \\
\hat{B}=\hat{b}_R^\dagger\hat{b}_L+\hat{b}_L^\dagger\hat{b}_R,\\ \\
\displaystyle{t(\hat{N})=-E_1-S_1-\frac{\hbar\eta^2U_0J_1}{\kappa^2+(\Delta_c-U_0J_0\hat{N})^2}},\\
\\
\displaystyle{f(\hat{N})=\frac{\hbar\eta^2}{\kappa}\arctan\left(\frac{\Delta_c-U_0J_0\hat{N}}{\kappa}\right),}
\end{array}
\end{equation}
and the coefficients are given in Eq.~(\ref{coeff2}).

Transforming the operators as
\begin{equation}
\left[\begin{array}{c} \hat{b}_+\\
\hat{b}_-\end{array}\right]=\frac{1}{\sqrt{2}}\left[\begin{array}{cc}1
& 1
\\ 1 & -1\end{array}\right]\left[\begin{array}{c} \hat{b}_L\\ \hat{b}_R\end{array}\right]
\end{equation}
the Hamiltonian is diagonalized
\begin{equation}
\hat{H}_{BH}=(E_0+S_0)\hat{N}+f(\hat{N})-t(\hat{N})(\hat{n}_+-\hat{n}_-),
\end{equation}
with the new number operators $\hat{n}_+=\hat{b}_+^\dagger\hat{b}_+$
and $\hat{n}_-=\hat{b}_-^\dagger\hat{b}_-$. For given atom number
$N$, there are $N+1$ equidistant energy levels separated by
$t(N)$. In this case we recover perfect Josephson oscillations with
an oscillation frequency $\Omega_{Rabi}(N)=2|t(N)|/\hbar$. For atomic
states with an uncertain number of atoms, the various $N$'s will induce
a collapse in the Josephson oscillations as the contributing terms move out of phase.
Such a collapse was discussed for the regular BEC DW in
Ref.~\cite{milburn}, where it derives from the atom-atom interaction
term. For moderate or large atom numbers it is appropriate to assume
an initial coherent atomic state
\begin{equation}
|\psi\rangle=e^{-\bar{N}/2}\sum_n\frac{\bar{N}^{n/2}}{\sqrt{n!}}|n,0\rangle.
\end{equation}
Here, $\bar{N}=\langle\hat{N}\rangle$ is the average number of atoms
and the state $|n,m\rangle$ gives the number $n$ of atoms in the
left well and $m$ atoms in the right well. For
$\bar{N}\rightarrow\infty$, the relative uncertainty $\delta
N=\Delta N/\bar{N}$, where $\Delta
N=\langle(\hat{N}-\bar{N})^2\rangle$, goes to zero representing the
"classical" (mean-field) limit. Depending on the particular
$N$-dependence of $t(N)$, the different Josephson oscillation
terms may return in phase causing the system to
revive~\cite{milburn}. The revival time $T_r$ can be estimated as
the time it takes for consecutive terms to build up a $2\pi$ phase
difference $T_r[\omega_J(\bar{N}+1)-\omega_J(\bar{N})]=2\pi$ giving
\begin{equation}
T_r\approx\pi\hbar\left(\left.\frac{\partial t(N)}{\partial
N}\right|_{N=\bar{N}}\right)^{-1}.
\end{equation}
The $N$-dependence of $t(N)$ is supposedly weak in the large atom
limit implying $T_r\rightarrow\infty$ as $N\rightarrow\infty$, as expected in the classical limit.
Collapse-revival patterns are a widespread phenomena in physics and
have especially been studied in the vibrational dynamics of
molecules~\cite{cr} and in cavity QED~\cite{cavityQED}, as direct
proofs of quantization of either the molecular vibrations or of the
electromagnetic field. In the BEC DW system the collapse-revivals
derive from the squared atom number operators, $\hat{n}_{L,R}^2$,
while in standard cavity QED it is typically an outcome of a square-root
dependence of photon numbers in the Jaynes-Cummings model~\cite{jc}, $\sqrt{\hat{n}}$. 
Here, the atom number dependence of $t(N)$ is presumably more complex and one thereby
expects a less pronounced collapse-revival structure.

\begin{figure}[ht]
\begin{center}
\includegraphics[width=8cm]{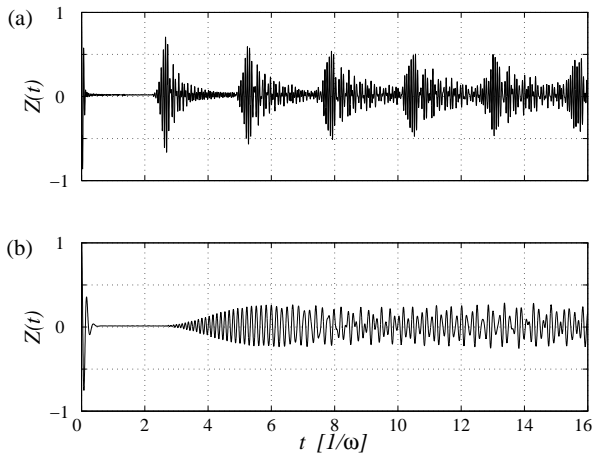}
\caption{Time evolution of the inversion $\mathcal{Z}(t)$ for an
initial coherent state in the right well with $\bar{N}=50$ (a) and
$\bar{N}=100$ (b). The common parameters are, $\kappa=2\pi\times1.3$
MHz, $\Delta_c=\kappa$, $\eta=2\kappa$, $U_0=\kappa/5$,
$\Delta_x\approx0.1$ $\mu m$, and $\omega=\kappa/500$. }
\label{fig10}
\end{center}
\end{figure}

In Fig.~\ref{fig10} we display the time evolution of the many-body
inversion
\begin{equation}
\mathcal{Z}_{MB}=\frac{\langle
\hat{n}_R\rangle-\langle\hat{n}_R\rangle}{\bar{N}}
\end{equation}
for an initial coherent atomic state in the right well with
$\bar{N}=50$ (a) and $\bar{N}=100$ (b). At short times, $t<0.3$,
a few Josephson oscillations persists before the collapse. The
collapse time is approximately the same for both examples, which is
a general property~\cite{cavityQED}. The first collapse period lasts until
$t\sim2.7$ for $\bar{N}=50$ and until $t\sim3.2$ for $\bar{N}=100$.
The $\bar{N}$-dependence in the revival times is as well a general
feature as argued above; large $\bar{N}$ values imply long revival
times~\cite{cavityQED}. In the upper plot (a), a sequence of
collapse-revivals appear, however less clear as time progresses. For
the lower plot (b), on the other hand, only a single collapse period
is visible. When the time-span of revivals begin to overlap, which
start to happen after roughly 1 ms in (a) and already after the
first revival in (b), super-revivals may occur due to higher order
interferences~\cite{frac}. Slight signatures of super-revivals can
be seen in (b) in the modulated oscillations after $t\sim10$. We
note that for the present results the time-spans are of the same
order as those for single Josephson oscillations in Fig.~\ref{fig10},
this derives from the much weaker pump amplitude $\eta$ in these
examples giving a lower barrier between the two wells, i.e. shorter
tunneling times.

As pointed out in the mean-field section, we have neglected effects
arising from atom-atom nonlinearities by letting $g=0$. The quadratic
terms in atom numbers are known to render
collapse-revivals~\cite{cr,milburn}, and with both nonlinearities
present (atom-atom and atom-field interactions) one would see a
competition between the two mechanisms. In general, revivals become
less frequent since they are only possible when both effects
simultaneously support revivals. We have verified these conclusions
numerically by including atom-atom interactions into our
calculations.

The analysis of this section relies on a static assumption, i.e. the
effective potential is considered independent on time. The
collapse-revival structure is solely an outcome of the uncertainty
of atom number and not of the instantaneous state of the atoms. The
full time-dependent many-body problem is certainly interesting, but
in this work we focus on general novel phenomena inherent in the
atom-field nonlinearity. Even though the results of these section
have been derived within some assumptions, we believe that the
general structure survives also in more rigorous analyzes. Such
approaches would for example reveal how the steady-state photon
number $n_{ss}=\langle\hat{n}_{ss}\rangle$ evolves in time. 
Nevertheless, it is believed that $n_{ss}$ encodes the properties of the atomic evolutions exactly as
it did in the previous section studying mean-field dynamics. That is,
one would expectedly find a similar collapse-revival pattern in
$n_{ss}$ as the one found in the atomic inversion.

It is important to note that the collapse-revival phenomenon is a pure quantum effect. By adding fluctuations around the means $\Psi(x)$ of the previous two sections one would typically encounter a collapse in the oscillations. However, the revivals depend strongly on the particular $N$-dependence of $t(N)$ and therefore random fluctuations around $\Psi(x)$ would not predict revivals. In other words, capturing the revival structure from some perturbed mean-field approach is only possible if one could carefully chose the perturbing fluctuations.

\section{Conclusions}
Nonlinear Josephson oscillations have been studied in a DW system of
ultracold bosonic atoms. Contrary to regular nonlinearity, arising
from atom-atom interaction in these types of systems, the
nonlinearity we considered derives from intrinsic interaction
between the atoms and a quantized cavity field. In particular, we
demonstrated the appearance of pseudo self-trapping, self-organization, as well as
collapse-revivals. In other words, the present work shows that these
phenomena are rather general and not restricted to only certain
kinds of nonlinearities.

The system parameters were chosen in agreement with current
experiments indicating that the phenomena should be experimentally
realizable. We furthermore showed or argued that the output
cavity field, proportional to $n_{ss}$, provides a direct handle of
the atomic dynamics. Since the output field is regarded as losses,
measurement of it is nondemolition by nature. Hence, the various
evolution regimes can be traced down without standard destructive
measurements such as time-of-flight or fluorescence detection.

\begin{acknowledgments}
JL acknowledges support from the MEC program (FIS2005-04627) and
VR/Vetenskapsr\aa det. We thank Anssi Collin, Maciej Lewenstein, Duncan O'Dell, Giovanna Morigi, and Jakob Reichel for insightful discussions.
\end{acknowledgments}

\end{document}